# Flexible scenario for heavy element research

D.Ibadullayev[a,b],*A.N.Polyakov[a], Yu.S.Tsyganov[a], A.A.Voinov[a]

[a]*Flerov Laboratory of Nuclear Reaction, Joint Institute for Nuclear Research, 141980 Dubna, Russia*

[b]*The Institute of Nuclear Physics, 050032 Almaty, Kazakhstan*

*e-mail:* ibadullayev@jinr.ru

**Abstract** – Detection system and software package coded in Builder C ++ are developed for the new analog spectrometer of the DGFRS-2 setup installed on the new DC-280 cyclotron. This software allows providing data acquisition using a 48x128 DSSSD (*Double Side Silicon Strip Detector*) detector and a multi-wire gaseous detector. The main feature of this software package is the use of flexible real-time algorithms to provide deep suppression of the background in reactions for the synthesis of superheavy elements, using the detection of correlations ER-α (Recoil-alpha particle). The block-diagram of analog spectrometer is presented.

## INTRODUCTION

The study of superheavy nuclei is arousing great interest in nuclear physics. Despite the fact that superheavy nuclei have very short lifetimes, various nuclear models can predict the existence of the so-called "island of stability" [1] - a region of nuclei around Fl, whose lifetime will be much longer compared to nuclei far from this region. The last currently known superheavy nuclei were discovered in experiments at the Dubna gas-filled recoil separator (DGFRS) FLNR JINR. These were fusion-evaporation reactions using actinide targets and $^{48}$Ca heavy ion beams accelerated at the U-400 accelerator. The intensity of these beams was about 1pμA.

For further research, a new accelerator DC-280 with an intensity of heavy ion beams up to 10 pμA was put into operation. When leaving the accelerator, heavy ion beams bombard a rotating target, as a result of which the reaction itself occurs. The first experimental setup of this Superheavy element factory is DGFRS-II, which is necessary for separating superheavy nuclei from heavy ion beams and other background particles formed as a result of the reaction [2].

## 1. THE DETECTION SYSTEM

The separated EVRs passed through a ΔE measurement system composed of multiwire proportional chamber placed in pentane at a pressure of 1.200 ±0.0017 Torr and were finally implanted in the detector module installed in the focal plane of the separator. The ΔE information is used to discriminate the particle that passed through the separator from α-particles or spontaneous fission (SF) of implanted EVRs. ΔE chamber is mounted 65 mm from another plane with -200 V potential in order to eliminate influence of space charge in the vicinity. A sensitive area of this chamber is about $230 \times 80$ mm$^2$. The focal-plane DSSD detector has 48 x 128 strips (1x2 mm$^2$ "pixels"; horizontal, vertical). The detector is used to measure energies of EVRs, sequential α-decays and spontaneous fission events. Their vertical position is determined from comparison of the charge signals measured at the top and the bottom of each strip. This detector is surrounded by side detectors without position sensitivity. Thus, detection efficiency for α-decay of implanted nuclei is increased up to about 95%. A set of similar 16 strip Si detector was mounted behind the detector array and operated in "veto" mode in order to eliminate signals from low-ionizing light particles, which could pass through the focal-plane detector (~330 μm) without being detected in the **ΔE** system. In the Fig.1a,b block-diagram for one electronics channel, CAMAC crate with ADP-16 units and view of mechanical housing of DSSD detector are shown. Data taking process is triggered with non-zero bit in the status register 1M unit which corresponds to any ***ADP-16 "L" TTL*** signal for seven modules. Three of them correspond to 48 strips of DSSD detector, whereas seven another's – to 64 side detectors. Gating Pa3n unit for reading ΔE signals is performed only when fast signal from any ADP for DSSD detector is existed. Duration of "GATE" signal is equal to 6 μs. Detection system has 2.8 μs dead time due to 8 cells ADP's FIFO, whereas regular dead time is about ~25 μs.

The main electronics units of the DGFRS-II setup are presented in the Table1. Their functions are mentioned in the row 3 of the Table in brief.

Fig. 1a. View of mechanical housing of DSSD detector (Micron Semiconductor, UK)

**Fig.1b Block-diagram of the analog spectrometer**

*Table 1 Main electronics unit*

| Number on Fig.2 | Unit name | Function (in brief) |
|---|---|---|
| 1 | Double Side Silicon Strip Detector | To measure ER, α, SF and other signals |
| 2 | Mesytec Mpr-64 | Charge sensitive preamplifier |
| 3 | ΔE low pressure pentane filled proportional chamber | To measure signal from charged particle coming from cyclotron DC-280 |
| 4 | Analog processor ADP-16 1M | 16-in, two scales shaping amplifier-analog multiplexer-analog to digital converter 13 bit first scale(<27 Mev), 12 bit second scale (< 270 Mev) |
| 5 | Pa3n | 3-in 1M unit to measure ΔE signals. 12 bit/channel |
| 6 | Ext-16 | 2M unit to create "STOP" TTL signal to stop irradiation process. |
| 7 | SU-212 2M | Shaping amplifier (FLNR,JINR design) to measure ΔE signals |
| 8 | 16 bit counter | 1M unit to measure target rotation speed, event rate of DSSD |
| 9 | XIA digital system | XIA Corporation digital spectrometer (autonomous) |
| 10 | Status 16 bit register | Start read-out process if are there any non-zero signals of "L" any ADP of front strips (48 strips) |

| | | Bit#16 1/0 beam off/on |
| | | Bits 1..3 – DSSD operation, 48 front strips; |
| | | 4..7 – SIDE detectors, 64 strips; |
| | | 8..15 – DSSD, 128 back strips. |
| 11 | Splitter unit 3M | 32-inunit to split signals from preamplifiers to digital and analog system(FLNR,JINR design) |
| 12 | 1M unit 6OR | 6-in logical TTL input signals to provide trigger TTL signal for gating of Pa3n unit |
| 13 | 1M unit PATS01 | 12 bit ADC to measure summary signal from all side detectors (JINR, FLNR design) |
| 14 | 1M unit AM-208 | 8- in analog multiplexer(JINR, FLNR design) |

## 2. THE BLOCK DIAGRAM OF THE MAIN PROCESS

In the Fig.2 the block diagram of the main data acquisition process is shown. These procedures, marked in grey are responsible for a real-time search of ER-α correlated sequences[4-6]. C++ program named YDA starts acquisition process when status register is greater than null in the first 7 bits position. Three bits are corresponded to focal plane 48 strips, whereas another four bits are corresponded to 64 strips of side detector. When data taking process takes place, each event specifies by 16 words of 16 bits. Thresholds of each ADP-16 CAMAC 1M unit are established by NAF function in the program booting stage. The detection system is described in [7] in details. A flexible scenario for real-time detection of ER-α sequence is described in [8]. For more details visualization programhas been developed. This program creates more than 300 histograms in off-line process. In the Fig. 3 examples of application YDA C++ code in heavy ion induced complete fusion reaction $^{242}$Pu + $^{48}$Ca → $^{287}$Fl + $3n$ and $^{243}$Am+$^{48}$Ca→$^{288}$Mc+$3n$ are shown. Note that shadows denote that the YDA C++ program for a short time created a beam stop in a real-time mode.

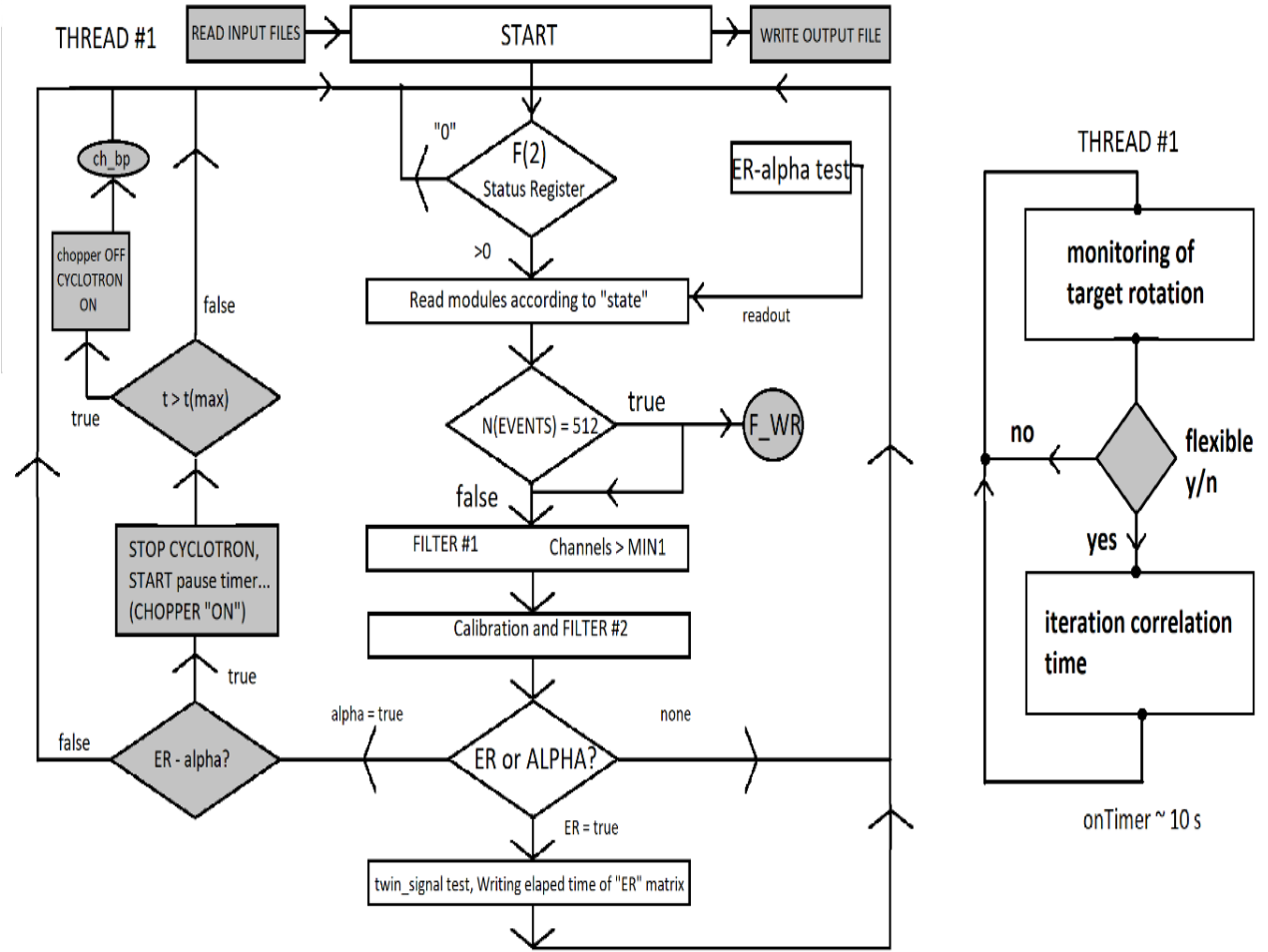

Fig.2 Block diagramm of YDA program main process. Main (CAMAC) process- priority *tpTimeCritical*, another Thread – *tpLowest* Priority of application – *RealTime*.

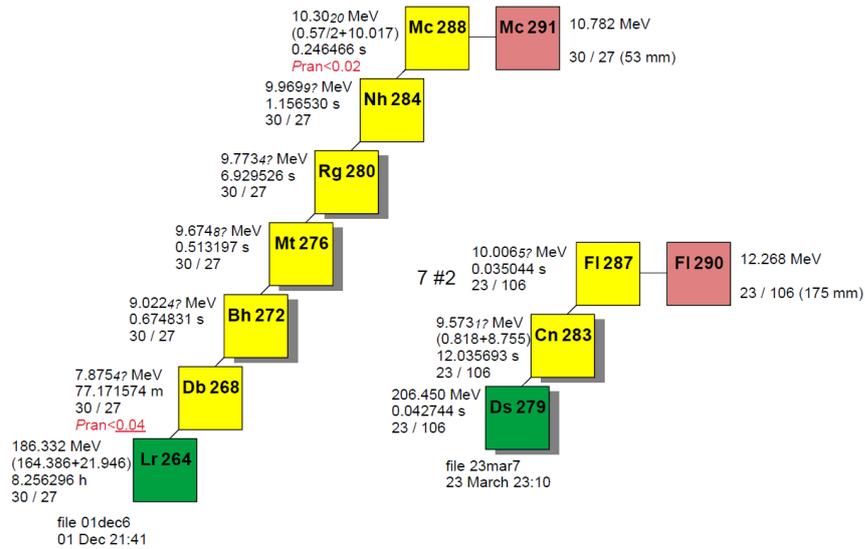

Fig.3. Examples of application of YDA C++ code in $^{48}$Ca induced complete fusion nuclear reactions

The "flexible" algorithm to search for ER-α correlation sequence in a real-time mode has been tested too. It means that initial value for recoil-alpha time interval was changed during the YDA code execution. The optimizing parameter of $N_b$(expectation per day) was in fact a number of beam stops due to random correlations. In the Fig. 4 the pre-setting parameter of $N_b$ was equal to 4 and correlated time was 20 s in a first approximation.

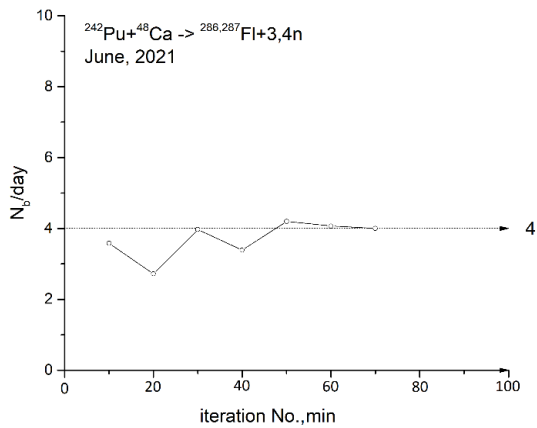

Fig. 4. Test of "flexible" algorithm. Pre-setting parameter of $N_b$ is equal to 4. Final ER-α time interval is about 40 s.

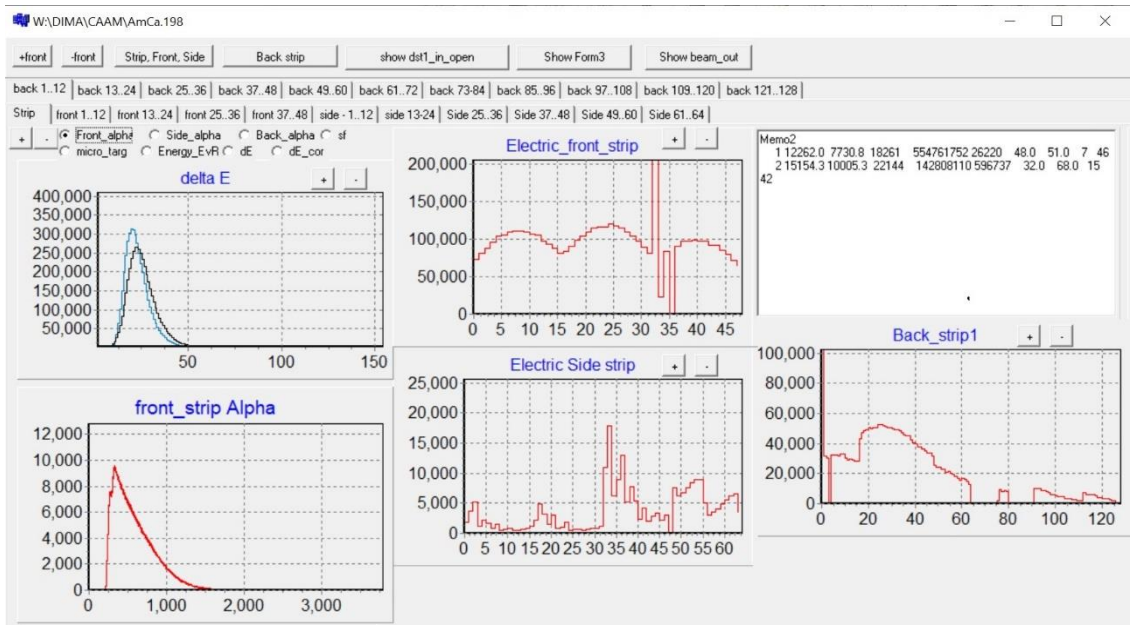

Fig. 5. Main interface view of Dastan#1 C++ program

## 3. A PROGRAM FOR PROCESSING THE RESULTS OF EXPERIMENTS ON DGFRS-II

This program is necessary for processing the results of experiments on the synthesis of superheavy nuclei at DGFRS-II. The visualization program allows to visually show the spectra of alpha events, fusion-evaporation residues and fission fragments from side, front and back strips. We can view the spectra on each strip separately. We can also observe the spectra of events from the time-of-flight camera. Another feature of this program is the search for correlation events and events out of beam, which makes it possible to find decay chains of superheavy nuclei. In Fig. 5 we can see the view of visualization program main interface.

### SUMMARY

DSSSD based detection system and C++ program package for new analog spectrometer of the DGFRS-2 setup are designed and successfully tested in $^{242}$Pu+$^{48}$Ca →*Fl and $^{243}$Am+$^{28}$Ca→*Mc complete fusion nuclear reactions. Complimentary the visualization program for files processing has been tested too. It is the first time a flexible algorithm for active correlation method has been applied in a test mode. An experiment $^{243}$Am +$^{48}$Ca is in progress now. We plan to apply this software in the near. We express my gratitude to Dmitry Solovyov and Maxim Shumeiko for their help in creating software packages. This paper is supported in part by the Ministry of Science and Higher Education of the Russian Federation through Grant No. 075-10-2020-117